\documentclass[reprint,aps,prl,showpacs,superscriptaddress]{revtex4-1}
\usepackage{graphicx,epsfig}
\usepackage{epstopdf}
\usepackage{amssymb,amsfonts,amsmath}

\begin{document}
\title{Dynamics of quantum turbulence of different spectra} 
\author{P. M. Walmsley}
\affiliation{School of Physics and Astronomy, The University of Manchester, Manchester M13 9PL, UK}
\author{D. E. Zmeev}
\affiliation{School of Physics and Astronomy, The University of Manchester, Manchester M13 9PL, UK}
\affiliation{Department of Physics, Lancaster University, Lancaster LA1 4YB, UK}
\author{F. Pakpour}
\affiliation{School of Physics and Astronomy, The University of Manchester, Manchester M13 9PL, UK}
\author{A. I. Golov}
\address{School of Physics and Astronomy, The University of Manchester, Manchester M13 9PL, UK}
\begin{abstract} Turbulence in a superfluid in the zero temperature limit consists of a dynamic tangle of quantized vortex filaments. Different types of turbulence are possible depending on the level of correlations in the orientation of vortex lines. We provide an overview of turbulence in superfluid $^4$He with a particular focus on recent experiments probing the decay of turbulence in the zero temperature regime below 0.5\,K. We describe extensive measurements of the vortex line density during the free decay of different types of turbulence: ultraquantum and quasiclassical turbulence in both stationary and rotating containers. The observed decays and the effective dissipation as a function of temperature are compared with theoretical models and numerical simulations.
\end{abstract}
\maketitle
\section{1. Introduction}
Superfluid helium is an ordered fluid with truly zero viscosity. Its local velocity is ${\bf v}_{\rm s} ({\bf r},t) = \frac{\hbar}{m_4}\nabla \theta$, where $\theta ({\bf r},t)$ is the macroscopically-coherent phase of the complex order parameter (here we refer to the common isotope $^4$He with the atomic mass $m_4$). 
As a result, the velocity circulation around any closed contour through the superfluid is always quantized in units of $\kappa \equiv h/m_4$. This allows stable topological defects --  quantized vortex lines, i.\,e. vortices with precisely one quantum of circulation around their filamentary ``cores'' just $a_0 \sim 1$\,\AA\ thick (inside which the order parameter is suppressed) \cite{Feynman1955,DonnellyBook}. Such a line can bend, move, and reconnect with another line when they come very close to each other. 
An array of vortex lines appears naturally as the equilibrium state of a superfluid rotating with the container \cite{HallVinen}. However, vortices can also form non-equilibrium tangles, known as quantum turbulence (QT) \cite{vinen08}, that possesses excess energy. As in any non-equilibrium system, the presence and structure of the vortex tangle is a result of the history of generation of vortices and flow in the liquid. 

In contrast to vortex lines, particle-like excitations (phonons and rotons) are in thermal equilibrium, so their density increases with temperature. These scatter off the cores of vortex lines exerting the ``mutual friction'' force on the superfluid. At temperatures below about 0.5\,K the excitations are too depleted to affect vortex dynamics and can be neglected. Between about 0.5 and 0.7\,K, they are nearly ballistic and can be treated as small dissipative perturbation that damps Kelvin waves (helical perturbations of the shape of vortex lines) \cite{hall60,DonnellyBook} of wavelengths smaller than $\eta_{\rm q}(T)$ which decreases with decreasing $T$ \cite{kozik08}. At higher temperatures excitations form a fluid of low viscosity (``the normal component'').  
The normal component can move at a velocity ${\bf v}_{\rm n}(\bf r)$ that is different from ${\bf v}_{\rm s}(\bf r)$. 
The presence of two co-existing but weakly interacting (they only interact in the presence of quantized vortices) fluids, superfluid and normal, both of which can be independently turbulent, is a unique feature of superfluid helium. At high temperatures, Kelvin waves of all wavelengths are damped. With decreasing temperature between some 0.7\,K and 0.5\,K, firstly larger and then smaller wavelengths of Kelvin waves become undamped \cite{kozik08, kozik08b}.  
Finally, in the $T=0$ limit, only Kelvin waves of extremely short wavelengths, $\eta_{\rm q0} \sim 100$\,\AA, remain dissipative due to the emission of phonons (i.\,e. density waves)  \cite{vinen01,KSPhonons}.

For $T \rightarrow 0$, the very idea of dissipation in a liquid devoid of viscosity and thermal excitations might sound counter-intuitive. 
One should realize that a turbulent superfluid is far from equilibrium; the energy of its flow depends on the configuration of vortex lines in the tangle. During the free decay, it evolves towards lower-energy configurations until no vortex lines are left. Vortex reconnections are vital events that allow the evolution of the topology of the tangle towards these lower-energy configurations with fewer vortex lines. In experimental containers, the later stage of the decay can be arrested by trapping  vortex lines in long-lived metastable configurations; however, this can only happen when not a tangle but a handful of lines remain \cite{awschalom84}. More generally, in various ordered condensed-matter systems, only in the absence of movable linear defects can one have truly non-dissipative states: e.\,g. persistent flow in superfluids devoid of vortex lines, persistent currents in superconductors devoid of flux lines, elastic deformation in single crystals devoid of dislocations, etc.

\begin{figure}
\begin{center}
\centerline{\includegraphics[width=.4\textwidth]{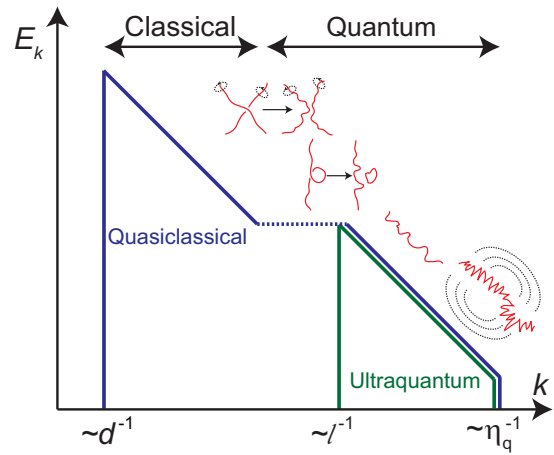}}
\caption{Schematic log-log energy spectra for quantum turbulence in the zero-temperature limit. At large {\it classical} length scales (small wavenumbers $k$) correlations between vortices allows classical Kolomogorov energy spectrum of flow coarse-grained over many vortex lines. At small {\it quantum} length scales (large $k$), comparable to the intervortex spacing $\ell$, the quantized nature of vorticity becomes important. The spectra for quasiclassical (blue lines) and ultraquantum turbulence (green lines) are shown. The cartoons indicate some of the different types of processes expected (from left to right): reconnections, self-reconnections and emission of vortex loops, Kelvin wave cascade, dissipation due to phonon emission from short wavelength Kelvin waves at the dissipative length scale $\eta_q$.\label{fig1}}
\end{center}
\end{figure}

Unlike classical fluids in which vorticity is a continuous field, the discrete nature of vortex lines in QT allows for the existence of a well-defined observable quantity -- the total length of vortex filaments per unit volume, $L$. This, in turn, introduces the important length scale -- the mean distance between neighboring vortex lines, $\ell \equiv L^{-1/2}$. At {\it classical} length scales much greater than $\ell$, one can neglect this discreteness and treat vorticity as a continuous field just like in classical fluids; on the other hand, at {\it quantum} length scales comparable to and below $\ell$, the local dynamics are dominated by the nearest quantized vortex lines. 
As in the case of other types of turbulence, the paradigm of quantum turbulence is that of a conservative cascade of energy towards the dissipative length scale. A schematic diagram of the energy spectrum of QT is shown in Fig.\,\ref{fig1}. At length scales greater than $\ell$, classical-like eddies form the Richardson (Kolmogorov) cascade. At much shorter length scales, a cascade is still possible but now due to the presence of discrete quantized vortex lines there is a {\it quantum cascade}. The particular processes resulting in the quantum cascade are thought to be various types of reconnections of vortex lines and non-linear interactions of Kelvin waves.

One can thus distinguish two limits of superfluid turbulence. At high temperatures ($1{\rm\,K} < T < 2.2$\,K), there are two interpenetrating fluids: the superfluid (harboring tangled vortex lines) and normal (the fluid of microscopic excitations) components. Historically, turbulence in superfluid helium was detected and studied in this high-temperature regime, and this continues to be an active area of research \cite{skrbek12}. In this regime, Kelvin waves of short wavelengths are strongly damped by their interaction with the normal component, hence the quantum cascade is not developed. The interesting new physics arises from the presence of two interacting turbulent components. 

Another special domain of interest, {\it which is the focus of this article}, is the limit of zero temperatures at which the excitations are either negligible or at best extremely depleted such that the mutual friction force on quantized vortices is only a small dissipative parameter. One thus deals with a single-component superfluid, in which the only dynamic degrees of freedom are the locations of the filaments of quantized vortex lines. These lines move with the local superfluid velocity which, thanks to the Biot-Savart theorem, depends on the positions of all vortex lines in the tangle. They bend and reconnect. Kelvin waves propagate without dissipation up to wavenumbers of order $\sim 10^7$\,cm$^{-1}$, above which damping is expected. One thus deals with a turbulent tangle whose dynamics spans many decades of length scales -- including a developed quantum cascade at length scales $\leq \ell$. While the temperatures $T \leq 0.5$\,K are rather easily achievable, several challenges, associated with the means of forcing and detecting QT, exist.  

From a more general point of view, vortex tangles are interesting for two reasons. Firstly, QT is another example of turbulence as a general phenomenon characterized by a multi-scale cascade of energy in systems with many degrees of freedom far from equilibrium. When forced by large-scale flow, QT has certain similarities with the classical hydrodynamic turbulence on large length scales but is very different at small, quantum, scales -- where there are similarities with wave turbulence. 

Secondly, vortex tangles are an example of interacting topological defects that can be found in many ordered systems (dislocations in solids, disclinations in liquid crystals, flux lines in superconductors, cosmic strings in the Universe, etc.). Many models of the dynamics of fast phase transitions into an ordered phase (when symmetry is spontaneously broken) assume creation of tangles of such defects. The dynamics of these defects are especially rich in superfluids with low dissipation because they become turbulent. Provided there is no large-scale bias, the nucleated tangles are expected to have no large-scale correlations; for the case of QT, such tangles would belong to the class of so-called ``random tangles'' which have no analogs in classical hydrodynamic turbulence.

In the following section we describe the different types of QT in more detail. We then focus on recent research in Manchester and describe the experimental techniques that have been developed for generating and detecting QT at low temperatures in section 3. In section 4, we review measurement of the free-decay of turbulence in a stationary container and include extensive new measurements that have been carried out to answer some outstanding issues from earlier work. Original research on the decay of turbulence in a rotating container is described in section 5. A new technique for forcing steady-state turbulence is presented in section 6 before we discuss the future outlook in the final section.

\section{2. Types of quantum turbulence}

Due to the existence of the quantum length scale $\ell$ (and hence of two different spectra of energy above and below the corresponding wavenumber as shown in Fig.\,\ref{fig1}), QT can have different dynamics depending on the wavenumber of forcing. This is different from classical turbulence, in which the spectra are always qualitatively similar. At the most basic level, there are two extreme types of QT: {\it quasiclassical} turbulence (forcing takes place at length scales much greater than $\ell$ and thus there is a substantial amount of energy in the classical part of spectrum) and {\it ultraquantum} turbulence (forcing occurs at length scales smaller than $\ell$ with a negligible amount of energy in the classical part of spectrum). For both cases, the dissipation of energy is through the motion of vortices, and it is assumed that the dissipation rate per unit mass is given by the heuristic relation 
\begin{equation}\label{eqnNu}
\dot{E}=-\nu(\kappa L)^2,
\end{equation}
where 
$\nu$ is an {\it effective kinematic viscosity} whose value is dependent on the particular type of QT.  Hence, measurements of $\nu$ can provide some insight into the underlying dynamics for different types of QT. 

While Eq.\,\ref{eqnNu} can be seen as a generalization of the classical-fluid analog for viscous losses, in which $\nu$ is the true kinematic viscosity (i.\,e. a material parameter independent of the type of flow) and $(\kappa L)^2$ stands for the effective total mean square vorticity, this is a dubious analogy. In essence, $\nu$ is a measure of the efficiency, for a given vortex line density $L$, of a particular type of vortex tangle at which the kinetic energy of flow is delivered down to the dissipative scale $\eta_{\rm q}$. One can arrive at this relation by considering vortex-vortex reconnection events (which are believed to be the key process of injecting energy into the quantum cascade towards the dissipative length scales) at the ultraquantum level: the number of elementary segments of radius $\sim \ell$ in a unit volume is $\sim L \ell^{-1}$, they are separated by the distance $\sim \ell$ and move with self-induced velocity $\sim \kappa \ell^{-1}$ (and hence their typical lifetime between reconnections is $\sim \ell(\kappa \ell^{-1})^{-1} = \kappa^{-1} \ell^2$), the average energy released upon each reconnection event is a fixed fraction of the energy of a segment $\sim \rho \kappa^2 \ell$. After combining all these one arrives at the energy flux per unit volume of order $\rho\dot{E} \sim -(L\ell^{-1})(\kappa\ell^{-2})(\rho\kappa^2\ell) = -\rho \kappa (\kappa L)^2$, i.\,e. $\nu \sim \kappa$ (according to Eq.\,\ref{eqnNu}). The value of the numerical prefactor, of course, depends on the microscopic structure of the particular vortex tangle (distribution of 
segments' shapes and sizes, correlations in velocities, relative angles of reconnections, etc.). In the $T=0$ limit, a broad range of length scales between $\ell \sim 10^2{\rm -}10^3$\,$\mu$m and $\eta_{\rm q} \sim 10^{-2}$\,$\mu$m are involved in the quantum cascade -- making the vortex length $L$ a fractal quantity \cite{svistunov95, kozik08, kozik08b}. In Eq.\,\ref{eqnNu} the full length $L$ (i.\,e. measured with the yardstick not greater than the short-scale fractal cut-off) should be used. This imposes an additional challenge on the calculation of the numerical prefactors in the above example. In fact, several different fractal regimes are actually predicted. The one with the most dramatic expansion of $L(\eta_{\rm q})$ with decreasing $\eta_{\rm q}(T)$ -- due to vortex-vortex reconnections -- for $\eta_{\rm q}$ just below $\ell$, while the ultimate mechanism due to non-linear Kelvin waves (without reconnections) at much shorter length scales $\ll \ell$ adds little to $L$ \cite{kozik08, kozik08b}. For practical  experimental purposes, it is hence sufficient to evaluate $L$ at length scales of order 1--10\,$\mu$m.

\subsection{Ultraquantum turbulence}
This is the simplest form of QT, where there is a nearly random tangle with no large-scale structures in the velocity field and thus it has no classical analog. The only characteristic length scale is the mean intervortex distance, $\ell\equiv L^{-1/2}$. Such a tangle can be created by forcing at length scales $< \ell$, i.\,e. by pumping energy into bending individual vortex lines. There are no large-scale correlations in the polarization of vortex lines, and hence the classical cascade is absent. The quantum energy is approximately proportional to the total length of vortex line of the tangle. After the forcing is no longer applied, the spatially-homogeneous turbulence is expected to decay according to
\begin{equation}\label{eqnUQ}
L=B\nu^{-1}t^{-1},
\end{equation}
where $B = \ln(\ell/a_0)/4\pi \simeq 1.2$. At these short length scales, the quantized nature of vorticity is dominant and non-classical processes are required to transfer energy to the short wavelength Kelvin waves where dissipation due to phonon emission is thought to be effective. Reconnections between vortices, and self-reconnections on individual vortex lines (which can lead to the emission of vortex rings) generate Kelvin waves on each vortex line (as shown in Fig.\,\ref{fig1}). Energy is thus transferred, with hardly any losses, to Kelvin waves of decreasing wavelength due to the non-linear Kelvin wave cascade, the precise details of which have been intensely debated during the last few years \cite{kozik09,kozik10,lvov10,boue11,sonin12}, until reaching the dissipative length scale. Thus, in the quantum regime of QT (at scales $\ll\ell$) the dominant excitations are expected to be ballistic vortex loops and Kelvin waves. At this stage, there is not yet any direct experimental evidence for the processes (Kelvin wave cascade, phonon emission, emission of vortex loops due to self-reconnections) involved in this scenario. However, ultraquantum turbulence has been observed experimentally. This was achieved by generating a beam of many small vortex rings and allowing them to coalesce into a nearly random tangle \cite{walmsley08a}.

\subsection{Quasiclassical turbulence}
When forcing turbulence on classical length scales ($\gg \ell$, by using a macroscopic propeller for example) then the classical spectrum is also excited. At these length scales, one expects that the coarse-grained vorticity becomes a continuous function and a classical-like description becomes possible (a Navier-Stokes-type equation governs the dynamics of the flow). The excitations of such a quasiclassical fluid are the usual hydrodynamic eddies and coherent structures due to bundles made up of many polarized vortex lines \cite{baggaley12a}. Thus, as well as the quantum energy, there is also a classical component of energy due to the kinetic energy of the large-scale eddies. If most of the energy is concentrated in the largest eddies (at the size of the container, $d$) then it can be shown \cite{StalpSkrbek} that the vortex length will decay as
\begin{equation}\label{eqnQC}
Ld^{-1} = (3C)^{3/2}(2\pi\kappa)^{-1}\nu^{-1/2}t^{-3/2},
\end{equation}
where $C=1.5$ \cite{sreeni} is the constant that appears in the classical Kolomogorov energy spectrum. This relation is the same as that observed in QT at higher temperatures and would also be expected in classical fluids \cite{skrbek12}.

In classical turbulence, the dissipative processes are heat losses through either shear or compressional flow, as well as the generation of sound waves. These mechanisms are only efficient at small length scales (large $k$) and high frequencies. The standard paradigm is that the kinetic energy of flow, generated at large length scales (small $k$), cascades with negligible losses to similar flows at ever decreasing length scales until the losses become substantial. A similar concept of a classical-like inertial cascade is applicable to QT. However, in the case of QT the sink of the classical energy is not the heat bath but the quantum cascade of the vortex tangle described above. How energy can be transferred from the classical cascade to the quantum cascade is another interesting aspect of QT in the zero-temperature limit. Kozik and Svistunov have suggested a scenario where various types of reconnections at the length scales in the vicinity of $\ell$ (including reconnections of vortex bundles, individual vortex lines, and self-reconnections) efficiently transfer energy to Kelvin waves \cite{kozik08,kozik08b,kozik09}, whereas L'vov, Nazarenko and Rudenko have developed a phenomenological model describing a gradual cross-over from the classical (Kolmogorov) cascade to the quantum (Kelvin wave) cascade \cite{lvov07,lvov08}. In Manchester, quasiclassical turbulence has been created in a series of experiments by rapidly bringing a rotating container to a halt \cite{walmsley07,walmsley08b}. We describe these experiments and compare the measured rates of dissipation to theoretical models in section~4.

\subsection{Rotating turbulence}
Our focus so far has been on the types of turbulence expected in a stationary container. However, the rate and efficiency of reconnections is expected to decrease with the alignment of neighboring vortex lines. This means that investigating vortex tangles that are polarized by applied rotation should provide further insight into the intrinsic mechanisms of dissipation of QT that we have already described. In classical fluids, turbulence in a rapidly rotating container becomes two-dimensional in nature and thus has very different dynamics when compared to isotropic turbulence \cite{Moisy,lambriben11}. Both the energy spectrum and free-decay laws are modified and an inverse cascade operates in parallel with the direct one. Rotation also permits bulk helical oscillations of the fluid known as inertial waves where the restoring force is provided by the Coriolis force. The frequency of these waves is less than $2\Omega$, and in a closed container resonant inertial modes occur at discrete frequencies that depend on the particular geometry \cite{lambriben11,bewley07}.

In a superfluid rotating at angular velocity $\Omega_0$, the equilibrium state is that with an array of parallel vortex lines with density $L_0 = 2\Omega_0/\kappa$. Vortex waves are possible and their nature will depend on the frequency of forcing, $\omega$ \cite{hall60,sonin87}. Low frequency excitations ($\omega<2\Omega$) are expected to be inertial waves \cite{hall60} due to the collective motion (such as compression and bending) of the vortex array, the same as in classical fluids. On the other hand, Kelvin waves on the individual vortices within the array are expected for excitations at higher frequencies \cite{hall60}. In the presence of an ordered vortex array, these can be further modified by its shear rigidity \cite{tkachenko} and collective pinning at container walls \cite{krusius1993}; however, we do not expect these issues to arise with turbulent vortex tangles. Thus, there is the potential to probe both classical and quantum regimes in rotating QT depending on the nature of the forcing. By gradually increasing the amplitude of forcing, there is also the opportunity to investigate the transition from an array of agitated vortices (highly polarized tangle) to turbulent vortex tangles (weakly polarized tangle).  At high temperature, rotating QT has been driven by applying an axial counterflow during steady rotation \cite{swanson83,tsubota04}, but new methods of forcing are required in the zero temperature limit \cite{walmsley12}. In section 6, we describe measurements on rotating quasiclassical and ultraquantum QT generated by mechanical agitation and pulsed charge injection respectively.
 
\section{3. Techniques}
\subsection{Detection of turbulence}
At temperatures above 1\,K, the vortex line density, $L$, is often measured by observing the attenuation of second sound. However, other techniques are required for probing turbulence at lower temperatures due to the vanishingly small normal component. In the Manchester experiments, we utilize injected electrons which were proved in the past to be effective probes of quantum turbulence \cite{careri60,gamota73,davis00}. In liquid helium, injected electrons self-localize in spherical cavities forming ``electron bubbles'' of radius $\sim 2$\,nm. These are often referred to as {\it negative ions}. Observing the attenuation of a beam of ions due to trapping and scattering from a turbulent tangle of vortices can be used to measure $L$ across a broad temperature range (0.05 -- 1.6\,K). At temperatures below 0.8\,K the electron bubbles are always attached to micron-sized vortex rings \cite{RR1964}. The interaction of these vortex rings with other vortex lines is geometrical, such that the effective scattering diameter is comparable to the rings diameter \cite{schwarz68}, making them particularly suitable as probes of turbulent vortex tangles. Another advantage is that the suitable radius (and hence, the value of the scattering diameter), typically $R = 1{\rm -}3$\,$\mu$m, of such charged rings can be prepared by passing them through a chosen potential difference. Last but not least, detecting the arrival of even small number of singly-charged vortex rings, through measuring their charge, is relatively straightforward. In a typical experiment, a $\sim 0.1$\,s pulse of electrons would be injected via field emission. The probe rings then traverse the main volume of the container before the current arriving at a detector electrode, shielded by a Frisch grid, is measured. Electric fields could be applied and used to manipulate injected charges by tuning the radius of charged vortex rings and controlling the amount of charge that enters the container.  Most experiments were performed in a cubic container with $d=4.5$\,cm that allowed $L$ to be measured along two orthogonal axes \cite{walmsley06} (see Fig.\,\ref{fig1}) but more recent measurements have used a channel of square cross-section with $d=1.27$\,cm. The minor drawback of this detection method is that it is an invasive procedure because the probe ions generate some additional turbulence, hence each tangle was usually only probed once before the experiment had to be repeated with the generation of a fresh vortex tangle.

\begin{figure}
\begin{center}
\centerline{\includegraphics[width=.5\textwidth]{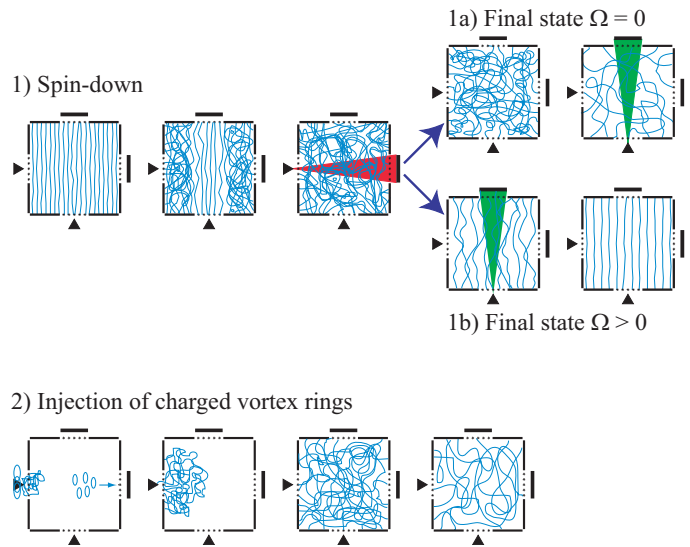}}
\caption{Cartoon of two different techniques of creating quantum turbulence: (1) spin-down to rest (a) and to finite $\Omega_0$ (b); (2) colliding charged vortex rings. For each type, the time evolution of the turbulence proceeds from left to right. In the early stages, the turbulence can be strongly inhomogeneous but eventually fills the entire container. Spin-down creates quasiclassical turbulence whereas short injection of vortex rings produces ultraquantum turbulence. The red (green) shaded areas indicate how $L$ can be probed in the horizontal (vertical) directions using a short pulse of injected charged vortex rings.\label{fig2}}
\end{center}
\end{figure}

\subsection{Generation of turbulence} 
\ An ideal experiment would have steady-state forcing of turbulence at a particular length scale. While this is not impossible, continuous forcing often complicates the process of measuring both $L$ and the injected energy. An alternative approach is to observe the free decay of turbulence. The energy flux down the cascade is limited by the slowest process, which is generally the breaking up of the largest eddies for classical cascade and reconnection-driven injection of energy into the quantum cascade. This is, of course, a non-steady-state. However, it can be considered as a quasi-steady state if all the vital processes that set the dissipation rate (and the instantaneous value of $L$) are fast compared to the slowest changes in forcing. In the Manchester experiments, the following two different means of forcing turbulence have been used. For both techniques, moving agitators were avoided -- because any friction can introduce excessive heating, which is detrimental in low-temperature experiments, although a novel and elegant technique for driving the motion of a grid has recently been developed \cite{zmeev13b}.

1. {\it Unsteady rotation.} Rotation has proved to be vitally important tool for generating QT in the zero-temperature limit. Changes in the angular velocity, $\Omega$, of the whole non-axially-symmetric container provides an effective way of creating large-scale turbulence. Another advantage is that steady rotation at angular velocity, $\Omega_0$, creates an array of rectilinear vortices of known density, $L_0 = 2\Omega_0/\kappa$, allowing the detection technique to be calibrated {\it in situ} as a function of temperature. The presence of a vortex array was confirmed by firing pulses along the axis of rotation and observing the distribution of arrival times because trapped electrons can slide quickly ($\sim 10$\,m\,s$^{-1}$) along vortex lines \cite{ostermeier75} compared to the relatively slow motion of charged vortex rings ($\sim 0.1$\,m\,s$^{-1}$). This has proved to be particularly useful for investigating the stability of a vortex array when perturbations due to an oscillatory component of rotation are applied \cite{walmsley12}. In the zero temperature limit, it typically takes $\simeq\,500$\,s after starting rotation for a vortex array to be stabilized.
For free-decay experiments, the turbulence was firstly prepared by either an impulsive spin-down from angular velocity $\Omega$ to rest or a sequence of alternating forward and backwards rotational agitations of amplitude $\Delta\Omega$ and zero mean. For steady-state forcing, continuous periodic modulations of the angular velocity of rotation (around either zero or non-zero mean value, if necessary) were the means of forcing turbulence at the chosen frequency and amplitude. 

2. {\it Electrostatic force on ions trapped on vortex cores.} This allows force to be directly exerted on segments of vortex lines and hence to pump energy both into small-scale perturbations of individual vortices and large-scale body force to the whole tangle. For instance, reconnections within a dense cloud of charged vortex rings injected during a short pulse produced a vortex tangle that is to a good approximation nearly random -- provided the total impulse (and hence quasiclassical energy of large-scale flow) transferred to the  tangle is relatively small. This allowed the dynamics of ultraquantum turbulence to be investigated, in which only the quantum part of the energy spectrum, and hence only the quantum cascade, is present. Because of this constraint on the relative magnitude of large-scale flow, only free-decay measurements upon a short ion injection at low temperatures were suitable for the studies of ultraquantum turbulence. Long and especially steady-state current injection resulted in most of the generated vorticity being in the form of entangled vortex lines ( and not individual vortex rings); the electrostatic force exerted on the trapped ions acted on the whole entrained superfluid, hence, producing quasiclassical flow. In addition, the process of entanglement caused by the reconnections between vortex rings can produce fluctuations in loop sizes and redistribute energy to both longer and shorter length scales \cite{baggaley12,baggaley12b}. For steady-state forcing, running a current of trapped ions of a chosen density through a space-filling vortex tangle, subject to the applied electric field, allowed the forcing of  large-scale flow (and occasionally small-scale flow in parallel as well -- at low temperatures). The nature (for instance, the drift velocity of ions) of the current flow permitted the distinction between the drift of the charge together with the tangle (quasiclassical large-scale flow) and the charge transport through small vortex loops that result from vortex reconnections at low temperatures (quantum cascade) -- as well as investigate these regimes.

The time evolution during the free decay of some of the different types of turbulence that are described in this article are shown schematically in Fig.\,\ref{fig2} (for the cubic container with $d=4.5$ cm). Firstly, impulsive spin-down to rest shown in Fig.\,\ref{fig2}(a), where $\Omega(t<0)=\Delta\Omega$ and $\Omega(t\geq 0)=0$ (in practice, these involved a rapid deceleration taking place over a few seconds) \cite{walmsley07,walmsley08b}. In addition, we have also used oscillatory AC-rotational agitation (where $\Omega(t<0)=\Delta\Omega \sin\omega t$ and $\Omega(t\geq 0)=0$). Secondly, spin-down to finite $\Omega=\Omega_0$ shown in Fig.\,\ref{fig2}(b) (where $\Omega(t<0)=\Omega_0 + \Delta\Omega$ and $\Omega(t\geq 0)=\Omega_0$). Thirdly, the turbulence created by a short ion injection (which is stopped at $t=0$), and collisions between vortex rings create a vortex tangle that spreads and fills the whole container (Fig.\,\ref{fig2}(c))\cite{walmsley08a}. In each case, the value of $L$ at a particular time due to the free decay of the turbulence was measured by observing the attenuation of a probe pulse of either charged vortex rings (for $T<0.7$\,K) or bare negative ions (for $T>0.7$\,K). 

\section{4. Decay of turbulence}
\begin{figure}
\centerline{\includegraphics[width=.5\textwidth]{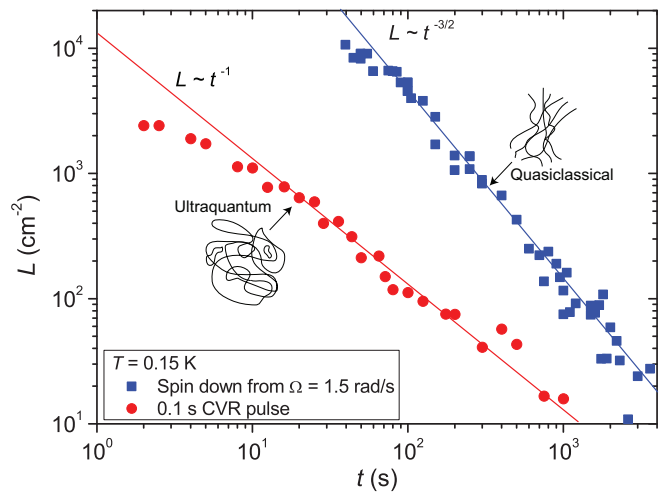}}
\caption{Comparison of the vortex line density during the free decay of turbulence created by spin-down and colliding charged vortex rings.\label{fig3}}
\end{figure}
The late-time behavior of the vortex line density at $T=0.15$\,K during the free decay of turbulence produced by an impulsive spin-down to rest and also by entangling a number of colliding vortex rings is shown in Fig.~\ref{fig3}. The results from these types of experiments have already been described in detail in earlier publications \cite{walmsley07,walmsley08a,walmsley08b}. In this article, we focus on the late-time decay where the turbulence is nearly homogeneous rather than the transient behavior during the early stages of the decay where the turbulence can be strongly inhomogeneous. The key salient feature is the different power-laws observed for these two methods of generating turbulence. The turbulence resulted from a  spin-down decays with $L(t)\sim t^{-3/2}$ which is indicative of quasiclassical turbulence of fixed spatial extent with $\nu = 0.003\kappa$ (Eq.\,\ref{eqnUQ}). It was found that $\nu$ was independent of $\Delta\Omega$. On the other hand, the turbulence created by colliding vortex rings decays with $L(t)\sim t^{-1}$ suggesting  ultraquantum turbulence, with no large-scale structure, and $\nu = 0.08\kappa$ (Eq.\,\ref{eqnQC}). In both cases, the values of $\nu$ did not depend on the direction along which $L$ was measured. Clearly, the rate of energy dissipation for both types of turbulence is finite in the zero temperature regime. The value of $\nu$ for ultraquantum turbulence is approximately a factor of 30 greater than for quasiclassical turbulence. This suggests that for the same value of $L$, energy is dissipated much faster when there are no flow patterns at large scale ($\gg \ell$, i.\,e. the classical part of the energy spectrum is absent) and there is only a quantum cascade.

\begin{figure}
\centerline{\includegraphics[width=.45\textwidth]{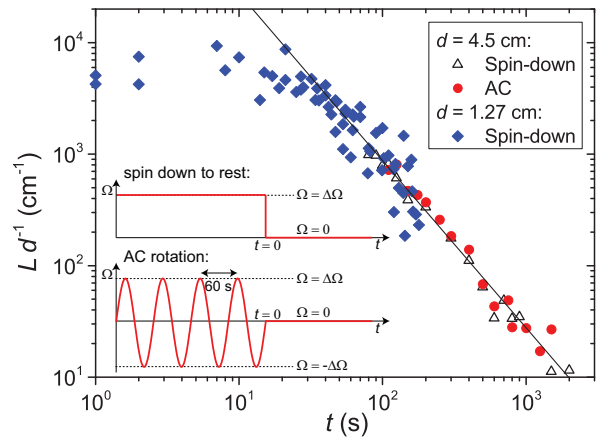}}
\caption{Quasiclassical turbulence created using different methods (impulsive spin-down to rest and a wave train of AC-rotational agitation with zero mean) and container sizes ($d=4.5$ cm and 1.27 cm). $\Delta\Omega=1.5$ rad s$^{-1}$ for all measurements. The solid line shows Eq.\,\ref{eqnQC} with $\nu=0.003\kappa$. Inset: Sketch of $\Omega(t)$ for the two methods of creating turbulence.\label{fig4}}
\end{figure}

However, following the original spin-down experiment, there were a number of potential issues identified regarding the interpretation of the experiment and the extracted value of $\nu$ at low temperatures \cite{golov09,vinen10}. The absence of normal fluid means that there was serious uncertainty regarding how quickly the energy of rotation was converted into turbulence and what the effect was on the turbulent decay due to any long-lived rotating state \cite{eltsov10}. Particularly open to questions was the assumption that the length scale of the large energy containing eddies was equivalent to the size of the container, $d$. At high temperature, the non-slip boundary condition for the normal fluid and the coupling due to mutual friction help mediate the decay of turbulence in the superfluid component. However, at low temperatures the boundary condition is less clear and depends on the surface interaction between vortices and the container walls. A recent experiment on $^3$He-B in a cylindrical container at low temperatures has shown that the response of the fluid following spin-down can be drastically altered by changing the effective boundary conditions \cite{walmsley11}. In order to address these issues, several further experiments have been performed which we describe here for the first time. Firstly, a series of grooves (0.5\,mm depth, 1\,mm width) were cut into the walls of the cubic container to see if they assisted with breaking up the large scale rotational flow. There was no discernible change on $L(t)$ following a spin-down. Secondly, the free decay following AC-rotational agitation and thirdly the impulsive spin-down of a container with a smaller value of $d=1.27$\,cm have been performed (whereas $d=4.5$\,cm for all the other measurements). The late-time decay of the turbulence produced in these experiments is shown in Fig.\,\ref{fig4}. The transient behavior in the early stages of the decay is very different when comparing the spin-down and AC-rotational agitation. This is because in the case of the spin-down the initial state is a polarized vortex array and $L$ increases immediately after the spin-down due to the transfer of energy from the largest length scale $\sim d$ to smaller scales as the Richardson cascade develops. However, the decay following AC-rotational agitation proceeds from an already developed turbulent state because the period of the AC rotation is much smaller than the time taken for a vortex array to form or decay. The key point is that the late-time decay does not depend on the creation method, and $L(t)$ scales as expected from Eq.\,\ref{eqnQC} if $d$ is taken to be the length scale for the largest energy-containing eddies in each case. Fitting each dataset to Eq.\,\ref{eqnQC} gives almost identical values of $\nu\simeq 0.003 \kappa$, thus this value seems likely to be representative of homogeneous quasiclassical turbulence. Further verification will require grid generated turbulence \cite{ihas08}, which is the benchmark for creating homogeneous turbulence in classical fluids.

\begin{figure}
\centerline{\includegraphics[width=.6\textwidth]{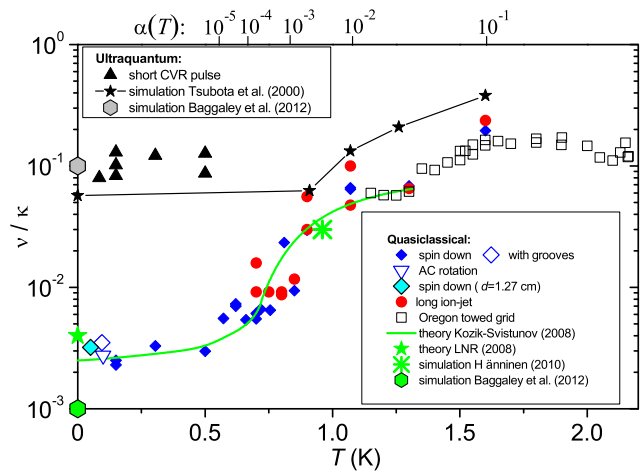}}
\caption{Effective viscosity versus temperature for quasiclassical ($L\sim t^{-3/2}$) and ultraquantum turbulence ($L \sim t^{-1}$). The upper axis shows the dissipative mutual friction parameter, $\alpha(T)$. The quasiclassical values are obtained from DC \cite{walmsley07,walmsley08b} and AC spin-down. The turbulence produced by ion jets is identical to that produced by rotation for $T\geq 0.7$\,K \cite{walmsley08a}. The high-temperature data for the Oregon towed grid experiments are shown for comparison \cite{stalp02,niemela05}. The ultraquantum turbulence was created by short CVR pulses \cite{walmsley08a}. Numerical simulations by Tsubota et al.\cite{tsubota00}, H\"{a}nnien \cite{eltsov10} and Baggaley et al. \cite{baggaley12} show good agreement with experiment for both types of turbulence as do the theoretical models of L'vov et al.\cite{lvov08} and Kozik and Svistunov \cite{kozik08b} which are also shown.\label{fig5}} 
\end{figure}

The dependence of $\nu$ on temperature for ultraquantum and quasiclassical turbulence is shown in Fig. \ref{fig5}. As our focus is on the low temperature behavior we omit values for most of the experiments that have been performed at high temperatures \cite{eltsov09,skrbek12}. In the high temperature regime, a whole range of techniques including counterflow, ultrasound, towed grids, ion jets and spin-down all have $\nu\sim 0.1\kappa$ for both ultraquantum and quasiclassical types of turbulence. However, as noted above at low temperatures the values of $\nu$ span a broad range $0.003\kappa - 0.1\kappa$. The ion injection technique appears to generate turbulence that is identical to spin-down when $T\geq 0.7$\,K, due to the force exerted by the ions trapped on the turbulent tangle helping to generate large-scale eddies of size $\simeq d$.

The value of $\nu\sim 0.1\kappa$ for ultraquantum turbulence is seemingly independent of temperature, although no single experiment has been able to span the whole temperature regime yet. On the other hand, $\nu$ for quasiclassical turbulence is temperature-dependent, indicating that, at low temperatures, there is a build-up of vorticity at scales close to $\ell$ when turbulence is forced at classical length scales. The phenomenological cross-over model developed by L'vov, Nazarenko and Rudenko \cite{lvov07,lvov08} can also account for the observed behaviour and the experimental value of $\nu$. Kozik and Svistunov \cite{kozik08b,kozik09} have used their reconnection-driven scenario to explain the temperature dependence of $\nu$ although a few fitting parameters are required. Nemirovskii has developed a model of the decay of an ultraquantum tangle based on the diffusion of vortex loops \cite{nemirovskii10} that showed certain qualitative agreement with our experimental data $L(t)$ for the decay of quasiclassical turbulence. Yet, even though a model of diffusion of a conserved vortex length does predict $L \propto t^{-3/2}$ decay, it contradicts our experimentally observed independence of late-time values $L(t)$ of the initial vortex density and scaling of $L(t) \propto d$ (which are, on the other hand, consistent with Eq.\,\ref{eqnQC}). Computer simulations of vortex tangles have been carried out by several groups. Recent simulations of decaying quasiclassical turbulence in a cube by H\"{a}nninen (but without surface pinning of vortices) \cite{eltsov10}, and of both ultraquantum and quasiclassical turbulence created by an anisotropic beam of vortex rings by Baggaley {\it et al.} \cite{baggaley12} are also in agreement with experiment. Thus, while identification of the two different structures of QT, ultraquantum and quasiclassical, has now been clearly observed along with the rate of energy dissipation, direct evidence for the predicted processes in the quantum regime (see section~2) is still required. Clearly, there is a diverse range of flows, associated with the motion, bending and reconnections of vortex lines, possible in the zero-temperature limit, making it ripe for further exploration.

\section{5. Turbulence in rotation}
\subsection{Free decay}
\begin{figure}
\centerline{\includegraphics[width=.5\textwidth]{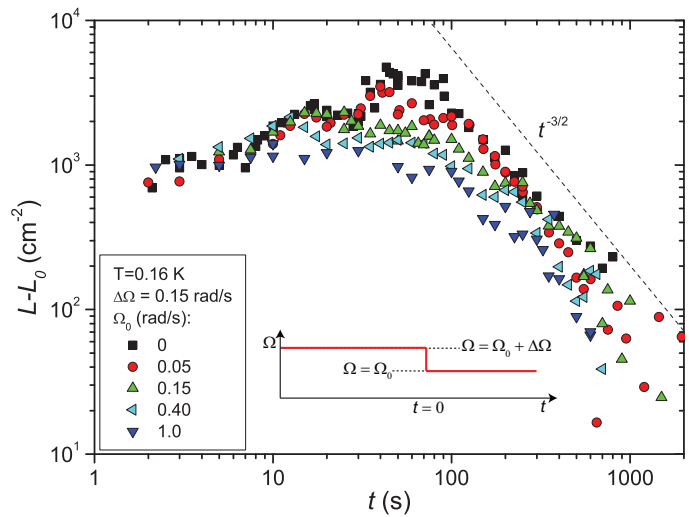}}
\caption{Time dependence of excess vortex line density during the free decay of rotating quasiclassical turbulence for different values of $\Omega$.  Inset: The turbulence was forced using spin-down with $\Delta\Omega=0.15$ rad/s. \label{fig6}}
\end{figure}

Quasiclassical rotating turbulence can be created by an impulsive spin-down to finite $\Omega_0$. As the steady final value of $\Omega_0$ increases, the effect of rotation on the decay of turbulence is expected to become more prominent. The excess vortex line density, $L-L_0$ (where $L_0=2\Omega_0/\kappa$ is the equilibrium value) for a series of spin-downs to different $\Omega_0$ but the same $\Delta\Omega=0.15$\,rad\,s$^{-1}$ (to assure that, in the reference frame co-rotating with the container in its final state, the amount of released energy is the same) is shown in Fig.\,\ref{fig6}. When $\Delta\Omega \geq \Omega_0$, then there is barely any difference compared to when $\Omega=0$. However, when $\Delta\Omega\ll\Omega_0$, the decay is markedly different. Firstly, the peak in $L-L_0$ during the transient early stages is suppressed indicating a reduced rate of growth of vorticity. Secondly, the late-time decay has a reduced slope which suggests that the free decay of vorticity is proceeding at a slower rate. These observations are in accordance with the observations on rotating turbulence in classical fluids \cite{Moisy,lambriben11} that  a greater amount of energy is stagnant at large length scales where it is stored in inertial waves and is slowly fed into the classical cascade towards smaller length scales -- whose flux is expected to vanish in the limit of high $\Omega_0$.

We have also attempted to generate a form of rotating ultraquantum turbulence by injecting a short pulse of ions on the rotational axis of the container. Some ions slide rapidly along vortex lines, generating Kelvin waves, and also charged vortex rings collide with vortex lines creating a turbulent tangle. In the presence of fast rotation, the excess vortex length did not rise as much initially and then decayed slower than in a stationary container, similar to the spin-down observations.

\subsection{Steady-state}
Steady-state rotating QT, rather than the freely decaying turbulence described above, can be forced by continuous oscillatory modulations of the angular velocity of rotating cubic container, $\Delta\Omega \sin(\omega t)$, where $\omega$ is the frequency of forcing (a similar forcing technique was recently used in the studies of inertial waves in a classical liquid \cite{boisson2012}). As well as measuring $L$, the connectivity of vortices can be probed in the axial direction by measuring the transport of trapped ions along vortex lines. The polarization can be controlled by varying the DC component of angular velocity. At small amplitude of forcing, there is a weakly perturbed vortex array with few reconnections but at high amplitudes of forcing the turbulence becomes more isotropic, making a fully-developed reconnecting tangle in which rectilinear vortex lines lose their identity. We summarize the main observations (see \cite{walmsley12} for details). Firstly, the transition from vortex array to bulk turbulence does require a finite level of forcing (albeit there is no observable threshold for the generation of boundary turbulence near a rough wall that tangles pinned vortex lines at very small forcing amplitude). Secondly, varying $\omega$ allowed resonances of inertial waves to be observed. The transition to bulk turbulence occurs at a lower amplitude of forcing when driven on an inertial wave resonance due to the enhanced fluid motion.

\begin{figure}[h]
\centerline{\includegraphics[width=.45\textwidth]{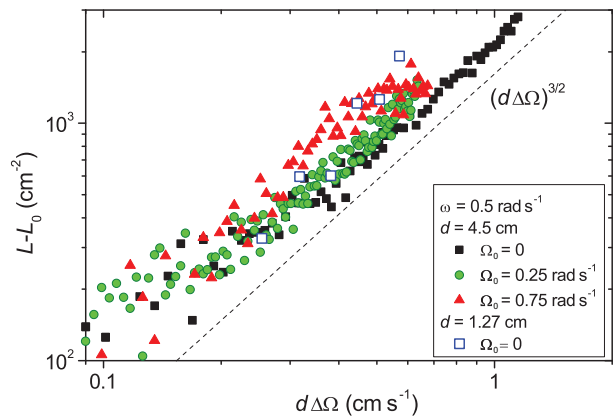}}
\caption{Dependence of the steady-state value of of excess vortex line density, $L-L_0$, on the amplitude of forcing velocity, $\sim d\Delta\Omega$ for QT that is driven by continuous AC oscillatory motion of two different containers ($d=4.5$ and 1.27\,cm) for both zero and non-zero mean rotation. The dotted line indicates $L\propto (d\Delta\Omega)^{3/2}$.
\label{fig8}}
\end{figure}

\section{6. Steady-state measurements}
Figure~\ref{fig8} shows the dependence of $L$ on the forcing amplitude, $\Delta\Omega$ for both zero and non-zero mean rotation. It appears that $L\sim \Delta\Omega^{3/2}$. The exponent $3/2$ can be explained by assuming that the fully-developed quasiclassical turbulence exerts a resistive torque $\propto \Delta\Omega^2$; then the time-averaged work of external forcing is $\propto \Delta\Omega^3$, that should be equal to the rate of dissipation, $\propto L^2$ (Eq.\,\ref{eqnNu}). As the torque exerted on the turbulent helium was not measured in these experiments, we do not know the energy pumped into the flow, and hence cannot extract the effective kinematic viscosity $\nu$. Future experiments, in which the energy flux will be monitored directly, will hopefully answer this question.

\section{7. Conclusions and future outlook}
QT in the limit of zero temperature provides a fascinating system for investigating many different facets of turbulence and vortex dynamics due to the quantized nature of vorticity. In recent years, our understanding of QT has improved substantially due to advances in computer simulations and analytical theory coupled with the development of new experimental techniques. Several types of turbulence, with different spectra, have been observed experimentally, depending on the length scale of forcing, including quasiclassical turbulence and ultraquantum turbulence. The rate of dissipation of both types has been measured, although new experiments that produce homogeneous and isotropic turbulence (such as grid turbulence) are desirable. The ability to rotate allows the polarization of turbulent tangles to be varied. Although these studies are still in their infancy, the decay of rotating quasiclassical QT is in accord with classical fluid behavior at large scales, including the observation of inertial wave resonances. Future experiments probing rotating QT at short length scales could provide the key to observing and characterizing the Kelvin wave cascade.

The techniques that have so far been used to probe QT in $^4$He at low temperatures are crude compared to modern studies of classical fluids that can visualize the velocity field and measure the energy spectrum. In a similar manner new techniques need to be developed to further our understanding of QT. The development of miniature sensors  (e.\,g. \cite{salort12}) that can detect fluctuations in velocity or pressure on length scales comparable to the intervortex spacing will allow the quantum regime to be probed. Visualization of vortex cores in quantum turbulence at high temperatures has become possible with the imaging of solid hydrogen particles trapped on vortex lines \cite{bewley08}. However, new techniques for visualizing vortex tangles in the zero temperature limit are needed. One promising route is to use helium excimer molecules that can be created {\it in situ} \cite{guo09}. It was found recently that these molecules become trapped on vortices below 0.2\,K \cite{zmeev13} raising the possibility of imaging vortex tangles through laser-induced fluorescence. This technique of mapping the field of {\it vorticity}, not velocity, opens advantages for both classical and quantum ranges of QT spectrum. 
It would allow phenomena across a broad range of length scales (from the classical 3d regime down to the 1d wave turbulence on individual vortices deep within the quantum regime) to be directly observed.

\begin{acknowledgments}
We are grateful to A.A. Levchenko, S. May, M. Sellers and S. Gillott for assistance with designing, constructing and improving the experiments. This research was funded by EPSRC (grant numbers GR/R94855, EP/H04762X and EP/I003738).
\end{acknowledgments}


\begin{thebibliography}{}
\bibitem{Feynman1955} Feynman RP (1955) Application of quantum mechanics to liquid helium. {\it Prog. Low Temp. Phys.} 1:17-53. 
\bibitem{DonnellyBook} Donnelly RJ (1991) Quantized vortices in helium II. Cambridge University Press, Cambridge.
\bibitem{HallVinen} Hall HE, Vinen WF (1955) Non-linear dissipative processes in liquid helium-II {\it Phil. Mag.} 46:546-548; (1956) The rotation of liquid helium-II. 1. Experiments on the propogation of 2nd sound in uniformly rotating helium-II {\it Proc. Roy. Soc. A} 238:204-214; (1956) The rotation of liquid helium-II. 2. The theory of mutual friction in uniformly rotating helium-II. {\it Proc. Roy. Soc. A} 238:215-234.
\bibitem{vinen08} Vinen WF (2008) An introduction to quantum turbulence. {\it Phil. Trans. R. Soc. A} 366:2925-2933.
\bibitem{hall60} Hall HE (1960) The rotation of liquid helium II. {\it Adv. Phys.} 9:89-146.
\bibitem{kozik08} Kozik E, Svistunov B (2008) Kolmogorov and Kelvin-wave cascades of superfluid turbulence at $T=0$: what lies between. {\it Phys. Rev. B} 77:060502(R).
\bibitem{vinen01} Vinen WF (2001) Decay of superfluid turbulence at a very low temperature: the radiation of sound from a Kelvin wave on a quantized vortex.  {\it Phys. Rev. B} 64:134520.
\bibitem{KSPhonons} Kozik E, Svistunov B (2005) Vortex-phonon interaction. {\it Phys. Rev. B} 72:172505.
\bibitem{awschalom84} Awschalom DD, Schwarz KW (1984) Observation of a remanent vortex-line density in superfluid helium. {\it Phys. Rev. Lett.} 52:49-52.
\bibitem{skrbek12} Skrbek L, Sreenivasan (2012) Developed quantum turbulence and its decay. {\it Phys. Fluids} 24:011301. 
\bibitem{kozik09} Kozik EV, Svistunov BV (2009) Theory of decay of superfluid turbulence in the low-temperature limit. {\it J. Low Temp. Phys.} 156:215-267.
\bibitem{kozik10} Kozik E, Svistunov B (2010) Geometric symmetries in superfluid vortex dynamics. {\it Phys. Rev. B} 82:140510(R).
\bibitem{lvov10} L'vov VS, Nazarenko S (2010) Spectrum of Kelvin-wave turbulence in superfluids. {\it JETP Letters} 91:428-434.
\bibitem{boue11} Bou\'{e} L et al. (2011) Exact solution for the energy spectrum of Kelvin-wave turbulence in superfluids. {\it Phys. Rev. B} 84:064516.
\bibitem{sonin12} Sonin EB (2012) Symmetry of Kelvin-wave dynamics and the Kelvin-wave cascade in the $T=0$ superfluid turbulence. {\it Phys. Rev. B} 85:104516.
\bibitem{walmsley08a} Walmsley PM, Golov AI (2008) Quantum and quasiclassical types of superfluid turbulence. {\it Phys. Rev. Lett.} 100:245301.
\bibitem{baggaley12a} Baggaley AW, Barenghi CF, Shukurov A, Sergeev YA (2012) Coherent vortex structures in quantum turbulence. {\it EPL} 98:26002.
\bibitem{StalpSkrbek} Skrbek L, Stalp SR (2000) On the decay of homogeneous isotropic turbulence. {\it Phys. Fluids} 12:1997-2019.
\bibitem{sreeni} Sreenivassan KR (1995) On the universality of the Kolmogorov constant. {\it Phys. Fluids} 7:2778. 
\bibitem{svistunov95} Svistunov B (1995) Superfluid turbulence in the low-temperature limit. {\it Phys. Rev. B} 52:3647.
\bibitem{kozik08b} Kozik E, Svistunov B (2008) Scanning superfluid turbulence cascade by its low-temperature cutoff. {\it Phys. Rev. Lett.} 100:195302.
\bibitem{lvov07} L'vov VS, Nazarenko SV, Rudenko O (2007) Bottleneck crossover between classical and quantum superfluid turbulence {\it Phys. Rev. B} 76:024520.
\bibitem{lvov08} L'vov VS, Nazarenko SV, Rudenko O (2008) Gradual eddy-wave crossover in superfluid turbulence. {\it J. Low Temp. Phys.} 153:140-161.
\bibitem{walmsley07} Walmsley PM, Golov AI, Hall HE, Levchenko, Vinen WF (2007) Dissipation of quantum turbulence in the zero temperature limit. {\it Phys. Rev. Lett.} 99:265302.
\bibitem{walmsley08b} Walmsley PM, Golov AI, Hall HE, Vinen WF, Levchenko AA (2008) Decay of turbulence generated by spin-down to rest in superfluid $^4$He. {\it J. Low Temp. Phys.} 153:127-139.
\bibitem{Moisy} Morize C, Moisy F (2006) Energy decay of rotating turbulence with confinement effects. {\it Phys. Fluids} 18:065107. 
\bibitem{lambriben11} Lambriben C, Cortet P, Moisy F, Maas LRM (2011) Excitation of inertial modes in a closed grid turbulence experiment under rotation. {\it Phys. Fluids} 23:015102.
\bibitem{bewley07} Bewley GP, Lathrop DP, Maas LRM, Sreenivasan KR (2007) Inertial waves in rotating grid turbulence. {\it Phys. Fluids} 19:071701.
\bibitem{sonin87} Sonin EB (1987) Vortex oscillations and hydrodynamics of rotating superfluids. {\it Rev. Mod. Phys.} 59:87-155.

\bibitem{tkachenko} Tkachenko VK (1966) Stability of vortex lattices. {\it Zh. Eksp. Teor. Fiz.} 50:1573  [{\it Sov. Phys. JETP} 23:1049].
\bibitem{krusius1993} Krusius M, Korhonen JS, Kondo Y, and Sonin EB (1993) Collective motion of quantized vortex lines in rotating superfluid $^3$He-B. {\it Phys. Rev. B} 47:15113. 
\bibitem{swanson83} Swanson CE, Barenghi CF, Donnelly RJ (1983) Rotation of a tangle of quantized vortex lines in He II. {\it Phys. Rev. Lett.} 50,190-193.
\bibitem{tsubota04} Tsubota M, Barenghi CF, Araki T, Mitani A (2004) Instability of vortex array and transitions to turbulence in rotating helium II. {\it Phys. Rev. B} 69:134515.
\bibitem{walmsley12} Walmsley PM, Golov AI (2012) Rotating quantum turbulence in superfluid $^4$He in the {\it T}=0 limit. {\it Phys. Rev. B} 86:060518(R).
\bibitem{careri60} Careri G, Scaramuzzi F, Thomson JO (1060) Heat flush and mobility of electric charges in liquid helium. {\it Il Nuovo Cimento} 18:957-966.
\bibitem{gamota73} Gamota G (1973) Creation of quantized vortex rings in superfluid helium. {\it Phys. Rev. Lett.} 31:517-520.
\bibitem{davis00} Davis SI, Hendry PC, McClintock PVE (2000) Decay of quantized vorticity in superfluid $^4$He at mK temperatures. {\it Physica B} 280:43-44.
\bibitem{RR1964} Rayfield GW, Reif F (1964) Quantized vortex rings in superfluid helium. {\it Phys. Rev.} 136:A1194-A1208.
\bibitem{schwarz68} Schwarz KW (1968) Interaction of quantized vortex rings with quantized vortex lines in rotating He II. {\it Phys. Rev.} 165:323-334.
\bibitem{walmsley06} Walmsley PM, Levchenko AA, Golov AI (2006) Experiments on the dynamics of vortices in superfluid $^4$He with no normal component.  {\it J. Low Temp. Phys.} 145:143-154.
\bibitem{zmeev13b} Zmeev DE (2013) A method for driving an oscillator at quasi-uniform velocity. {\it arXiv:1309.3141} [DOI: 10.1007/s10909-013-0942-2].
\bibitem{ostermeier75} Ostermeier RM, Glaberson WI (1975) Motion of ions trapped on vortices in He II. {\it Phys. Rev. Lett.} 35:241-244.
\bibitem{baggaley12} Baggaley AW, Barenghi CF, Sergeev YA (2012) Quasiclassical and ultraquantum decay of superfluid turbulence. {\it Phys. Rev. B} 85:060501(R).
\bibitem{baggaley12b} Baggaley AW, Barenghi CF, Sergeev YA (2012) Three-dimensional inverse energy cascade induced by vortex reconnections. {\it arXiv:}1208.5204.
\bibitem{golov09} Golov AI, Walmsley PM (2009) Homogeneous turbulence in superfluid $^4$He in the low-temperature limit: experimental progress. {\it J. Low Temp. Phys.} 156:51-70.
\bibitem{vinen10} Vinen WF (2010) Quantum turbulence: achievements and challenges. {\it J. Low Temp. Phys.} 161:419-444.
\bibitem{eltsov10} Eltsov VB, et al. (2010) Vortex formation and annihilation in rotating superfluid $^3$He-B at low temperatures. {\it Phys. Rev.} 161:474-508.
\bibitem{walmsley11} Walmsley PM et al. (2011) Turbulent vortex flow responses at the AB interface in rotating superfluid $^3$He-B. {\it Phys. Rev. B} 84:184532.
\bibitem{ihas08} Ihas GG, Labbe G, Liu SC, Thompson KJ (2008) Preliminary measurements on grid turbulence in liquid $^4$He. {\it J. Low Temp. Phys.} 150:384-393.
\bibitem{eltsov09} Eltsov VB et al. (2009) Turbulent dynamics in rotating helium superfluids, In {\it Progress in Low Temperature Physics:Quantum Turbulence}, eds. Halperin WP and Tsubota M, volume 16, pp. 45-146.
\bibitem{nemirovskii10} Nemirovskii SK (2010) Diffusion of inhomogeneous vortex tangle and decay of superfluid turbulence. {\it Phys. Rev. B} 81:064512.
\bibitem{stalp02} Stalp SR, Niemela JJ, Vinen WF, Donnelly RJ (2002) Dissipation of grid turbulence in helium II. {\it Physics of Fluids} 14:1377-1379.
\bibitem{niemela05} Niemela JJ, Sreenivasan KR, Donnelly RJ (2005) Grid generated turbulence in helium II. {\it J. Low Temp. Phys.} 138:537-542.
\bibitem{tsubota00} Tsubota M, Araki T, Nemirovskii SK (2000) Dynamics of vortex tangle without mutual friction in superfluid $^4$He. {\it Phys. Rev. B} 62:11751-11762.
\bibitem{boisson2012}  Boisson J, et al. (2012) Inertial waves and modes excited by the libration of a rotating cube. {\it Phys. Fluids} 24:076602. 
\bibitem{salort12} Salort J, Monfardini A, Roche PE (2012) Cantilever anemometer based on a superconducting micro-resonator: application to superfluid turbulence. {\it Rev. Sci. Instrum.} 83:125002.
\bibitem{bewley08} Bewley GP, Paoletti MS, Sreenivasan KR, Lathrop DP (2008) Characterization of reconnecting vortices in superfluid helium. {\it Proc. Nat. Acad. Sci.} 105:13707-13710.
\bibitem{guo09} Guo W, et al. (2009) Metastable helium molecules as tracers in superfluid $^4$He. {\it Phys. Rev. Lett.} 102:235301.
\bibitem{zmeev13} Zmeev DE, et al. (2013) Excimers He$_2^*$ as tracers of quantum turbulence in $^4$He in the {\it T}=0 limit. {\it Phys. Rev. Lett.} 110:175303. 
\end{thebibliography}
\end{document}